\documentclass[%
 prb,
 amsmath,amssymb,
 reprint,%
 unsortedaddress,
 superscriptaddress,
]{revtex4-2}
\usepackage[makeroom]{cancel}
\usepackage{bm}
\usepackage{amsmath}
\usepackage{mathtools}
\usepackage{amsfonts}
\usepackage{siunitx}
\usepackage{graphicx}

\newcommand{\be}{\begin{equation}}
\newcommand{\ee}{\end{equation}}

\usepackage{hyperref}
\usepackage{adjustbox}
\usepackage[dvipsnames]{xcolor}

\begin{document}

\title{Micromagnetic study of inertial spin waves in ferromagnetic nanodots}

\author{Massimiliano d'Aquino}
\affiliation{Department of Electrical Engineering and ICT, University of Naples Federico II, Naples, Italy}
\email{mdaquino@unina.it}
\author{Salvatore Perna}
\affiliation{Department of Electrical Engineering and ICT, University of Naples Federico II, Naples, Italy}
\author{Matteo Pancaldi}
\affiliation{Department of Molecular Sciences and Nanosystems, Ca’ Foscari University of Venice, 30172 Venice, Italy}
\affiliation{Elettra-Sincrotrone Trieste S.C.p.A., 34149 Basovizza, Trieste, Italy}
\author{Riccardo Hertel}
\affiliation{Universit{\'e} de Strasbourg, CNRS, Institut de Physique et Chimie des Mat{\'e}riaux de Strasbourg, F-67000 Strasbourg, France}
\author{Stefano Bonetti}
\affiliation{Department of Molecular Sciences and Nanosystems, Ca’ Foscari University of Venice, 30172 Venice, Italy}
\author{Claudio Serpico}
\affiliation{Department of Electrical Engineering and ICT, University of Naples Federico II, Naples, Italy}

\date{\today}
\begin{abstract}
Here we report the possibility to excite ultra-short spin waves in ferromagnetic thin-films by using time-harmonic electromagnetic fields with terahertz frequency. Such ultra-fast excitation requires to include inertial effects in the description of magnetization dynamics. In this respect, we consider the inertial Landau-Lifshitz-Gilbert (iLLG) equation and develop analytical theory for exchange-dominated inertial spin waves. The theory predicts a finite limit for inertial spin wave propagation velocity, as well as spin wave spatial decay and lifetime as function of material parameters. Then, guided by the theory, we perform numerical micromagnetic simulations that demonstrate the excitation of ultra-short inertial spin waves (20 nm long) propagating at finite speed in a confined magnetic nanodot. The results are in agreement with the theory and provide the order of magnitude of quantities  observable in realistic ultra-fast dynamics experiments.
\end{abstract}
\maketitle

\section{Introduction}

The study of ultra-fast magnetization processes is a central issue in spin dynamics for its potential application to future generations of nanomagnetic and spintronic devices\cite{dieny2020opportunities}. In the last decades, after the pioneering experiment\cite{beaurepaire1996ultrafast} revealing subpicosecond spin dynamics, the investigation of ultra-fast magnetization processes has increasingly attracted the attention of many research groups stimulating the production of considerable research\cite{koopmans2000ultrafast,stamm2007femtosecond,stanciu2007all,kimel2009inertia,kirilyuk2010ultrafast,lambert2014all,dornes2019ultrafast,hudl2019nonlinear}.  

Recently, the direct
detection of spin nutation in ferromagnets achieved experimentally\cite{neeraj2021inertial,Unikandanunni_PRL_2022} in the terahertz range has confirmed the presence of inertial effects in magnetization dynamics which were theoretically predicted\cite{ciornei2011magnetization,olive2012beyond,Mondal2017} several years ago. In the past decades, nutation-type
magnetization motions in nanomagnets were
also studied theoretically within the classical
dynamics occurring at gigahertz frequencies\cite{serpico_quasiperiodic_2004}. Besides its fundamental implications for the physics of magnetism, terahertz spin nutation opens the way to study possible exploitation of novel ultra-fast regimes for technological applications such as, for instance, 
% magnetization    switching\cite{neeraj2022inertial}.
%
ballistic magnetization  switching\cite{bauer2000switching,bertotti_geometrical_2003,daquino_numerical_2004,devolder2006precessional} driven by strong picosecond field pulses into the inertial regime\cite{neeraj2022inertial,Winter2022}.

From the theoretical point of view, inertial magnetization dynamics can be modeled by augmenting the classical Landau-Lifshitz-Gilbert (LLG) precessional dynamics with a torque term taking into account angular momentum relaxation\cite{ciornei2011magnetization}, which is able to explain the observed nutation dynamics\cite{neeraj2021inertial} in homogeneously-magnetized samples. When spatial changes of magnetization are allowed in magnetic systems of nano- and micro-scale, the issue of the emergence of inertial spin waves oscillating at terahertz frequency arises. In this respect, very recently a number of theoretical studies have been performed to characterize nutation spin waves\cite{Kikuchi2015,Giordano2020,Makhfudz2020nutation,lomonosov2021anatomy,cherkasskii2021dispersion,mondal2022inertial,Titov2022,Gareeva2023}. These interesting studies are mostly concerned with the analysis of spin waves propagation in infinite media. However, the realization of nanoscale magnetic devices such as, for instance, memories or computing units, does intrinsically involve confined nanostructures. 

In this paper, we investigate, by using full micromagnetic simulations of inertial LLG  (iLLG) dyanmics, the excitation of ultra-short inertial spin waves in a confined ferromagnetic nanodot under the action of terahertz fields. We first derive suitable dispersion relation under the assumption of exhange-dominated spin waves, then perform full micromagnetic ac steady-state analysis of magnetization response to assess the onset of nutation resonance at certain terahertz frequency. By choosing an excitation frequency larger than such nutation resonance, we demonstrate the possibility to excite short-wavelength (i.e. 20 nanometers long) nutation spin waves. Finally, we simulate realistic time-domain magnetization processes driven by subpicosecond excitation which reveal the finite speed propagation of these ultra-short inertial spin waves in possible experiments.

\section{Model of inertial spin wave dynamics}

Magnetization dynamics is described by the iLLG equation, which can be written in normalized form as follows\cite{ciornei2011magnetization,wegrowe2012magnetization,neeraj2021inertial,neeraj2022inertial}:
\begin{equation}\label{eq:iLLG}
\frac{\partial \bm m}{\partial t}= -\bm m\times \left(\bm h_\mathrm{eff} -\alpha \frac{\partial \bm m}{\partial t} - \xi \frac{\partial^2 \bm m}{\partial t^2} \right) \,,
\end{equation}
where magnetization is expressed by the unit-vector $\bm m(\bm r, t)$ normalized by the saturation magnetization $M_s$, time is measured in units of $(\gamma M_s)^{-1}$ ($\gamma$ is the absolute value of the gyromagnetic ratio), $\bm h_\mathrm{eff}$ is the micromagnetic effective field (also normalized by $M_s$) which includes contributions arising from different interaction (exchange, anisotropy, magnetostatic, Zeeman) terms in the free energy, $\alpha$ is the Gilbert damping parameter, and $\xi$ measures the strength of inertial effects. We remark that this dimensionless quantity can be written as $\xi=(\gamma M_s \tau)^2$ and, in this respect, it determines the physical time-scale $\tau$ of inertial effects, which according to previous studies\cite{ciornei2011magnetization, neeraj2021inertial,neeraj2022inertial} has the order of magnitude of fractions of picosecond (meaning $\xi\sim 10^{-2})$. Thus, the inertia in magnetization dynamics is controlled by a small quantity comparable with usual Gilbert damping $\alpha\sim 10^{-2}$.

Although the inertial effects represent a small term in eq. \eqref{eq:iLLG}, the iLLG dynamics is significantly different from classical precessional dynamics in that emergence of ultra-fast nutation appears at terahertz frequencies, which opens the possibility to access novel dynamical magnetization regimes. In the sequel, we will focus the attention on the possibility to drive the excitation of ultra-short spin waves in ferromagnetic thin-films.  

%\subsection{Linear spin waves in the inertial regime}

To this end, we will first consider the idealized situation of indefinite magnetic thin-film and derive the spin wave dispersion relation for small-amplitude spin waves in the exchange-dominated case.
For the sake of simplicity, we consider small magnetization oscillations around a spatially-uniform in-plane equilibrium $\bm m_0$ such as that obtainable by saturating the thin-film with a static external field $\bm h_a$. By posing $\bm m(\bm r,t)=\bm m_0+\delta \bm m(\bm r,t)$ and linearizing eq. \eqref{eq:iLLG} around $\bm m_0$, one has:
\begin{equation}\label{eq:linear iLLG}
\frac{\partial \delta \bm m}{\partial t}= -\bm m_0\times \left(\delta\bm h_\mathrm{eff} - h_0 \delta\bm m -\alpha \frac{\partial \delta \bm m}{\partial t} - \xi \frac{\partial^2 \delta \bm m}{\partial t^2} \right) \,,
\end{equation}
where $\delta\bm m \cdot \bm m_0=0$ (at first-order), $\delta \bm h_\mathrm{eff}$ only includes linear terms in the magnetization deviation $\delta\bm m$ (typically exchange, magnetostatics, uniaxial anisotropy) and $h_0=h_a+\kappa_\mathrm{an}>0$ ($\kappa_\mathrm{an}$ is the normalized uniaxial anisotropy constant, if applicable). The assumption of indefinite thin-film allows to consider description of magnetization perturbation $\delta \bm m$ in terms of plane waves via Fourier approach. Since we are interested in the study of short wavelength spin waves oscillating at terahertz frequency, we neglect magnetostatics in eq. \eqref{eq:linear iLLG} assuming $\delta \bm h_\mathrm{eff}=l_\mathrm{ex}^2\nabla^2 \delta\bm m$ (with $l_\mathrm{ex}=\sqrt{2A/(\mu_0 M_s^2)}$ and $A$ being the exchange constant of the material). 
We consider an in-plane magnetization equilibrium $\bm m_0=\bm e_x$ ($\bm e_x$ is the cartesian unit-vector along the $x$ axis) and consider for simplicity magnetization deviations $\delta \bm m(x,t)$ with spatial changes occurring only along the $x$ axis. In this situation, one can perform two-dimensional Fourier transform of eq.\eqref{eq:linear iLLG} leading to:
\begin{equation}\label{eq:linear iLLG Fourier}
    i\omega \delta\hat{\bm m} = -\bm e_x \times (-k^2 l_\mathrm{ex}^2 - h_0 - i\alpha\omega +\xi \omega^2)\delta\hat{\bm m} \,
\end{equation}
where $\delta\hat{\bm m}=\int\!\!\!\int_{\mathbb{R}^2} \delta\bm m(x,t) e^{i(k x-\omega t)}\, dx \,dt$. 
By expressing $\delta\hat{\bm m}=\delta\hat{m}_y \bm e_y + \delta\hat{m}_z \bm e_z$ and introducing the notation $\hat{\psi}=\delta\hat{m}_y+i \delta\hat{m}_z$, the latter equation becomes:
\begin{equation}\label{eq:linear iLLG psi}
    (-\omega-k^2 l_\mathrm{ex}^2-h_0 -i\alpha\omega +\xi \omega^2)\hat{\psi}=0 \,.
\end{equation}

\begin{figure}[t]
    \centering
    \includegraphics[width=0.5\textwidth]{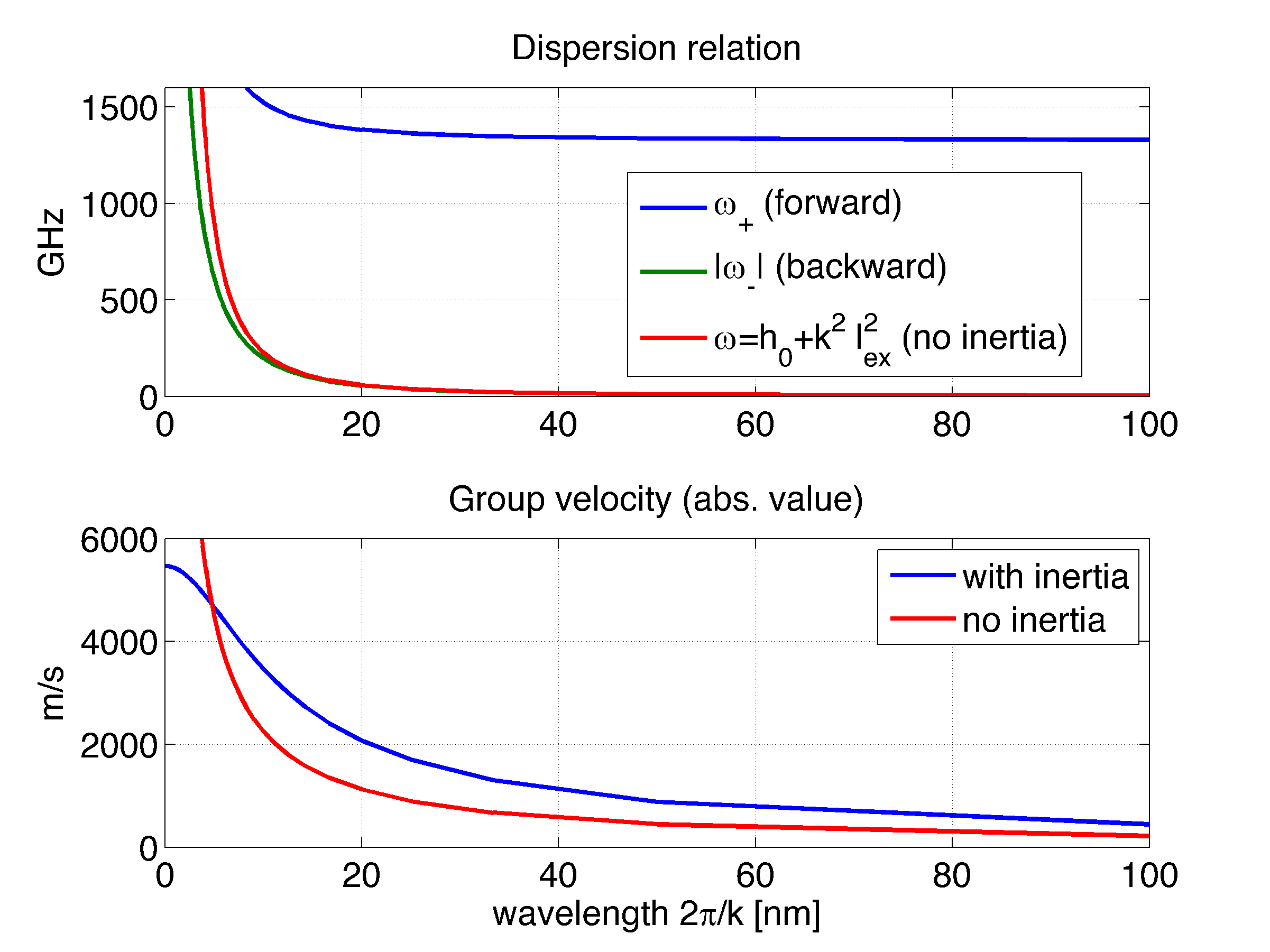}
    \caption{Dispersion relation for spin waves. Top panel reports $|\omega_\pm |$ according to eq.\eqref{eq:iLLG dispersion relation} along with the classical exchange-dominated spin wave dispersion relations. The bottom panel reports the associated group velocities. The value of parameters are $\gamma=2.211\times 10^5$ m$\,$A$^{-1}\,$s$^{-1}$, $\mu_0 M_s=1.6$ T, $A=13$ pJ/m ($l_\mathrm{ex}=3.57$ nm), $\tau=0.653$ ps $(\xi=0.0338)$ and $h_0=0.0625$. Notice the approach to the limiting value $v_g(k\rightarrow\infty)=l_\mathrm{ex}/\tau\approx 5470$ m/s in the short wavelength limit.}
    \label{fig:dispersion relation}
\end{figure}

Equation \eqref{eq:linear iLLG psi}, in the lossless limit $\alpha=0$, admits nontrivial solutions when the quantity in parenthesis vanishes:
\begin{equation}\label{eq:linear iLLG nontrivial}
    \omega^2-\frac{\omega}{\xi}- \frac{k^2 l_\mathrm{ex}^2}{\xi}-\frac{h_0}{\xi}=0 \,.
\end{equation}
The solution of eq.\eqref{eq:linear iLLG nontrivial} yields the dispersion relation (the angular frequency $\omega$ is measured in units of $\gamma M_s$):
\begin{equation}\label{eq:iLLG dispersion relation}
    % \omega_\pm=\frac{1}{2\xi} \pm \sqrt{\frac{1}{4\xi^2}+\frac{h_0+k^2 l_\mathrm{ex}^2}{\xi}} \,,
        \omega_\pm=\frac{1}{2\xi}\left(1 \pm \sqrt{1+4(h_0+k^2 l_\mathrm{ex}^2)\xi} \right) \,,
\end{equation}
which apparently is composed of two different branches (we remark that choosing $\hat{\psi}=\delta\hat{m}_y-i \delta\hat{m}_z$ yields the opposite of eq.\eqref{eq:iLLG dispersion relation}). The lower frequency branch $|\omega_-|$ represents the deviation from the classical exchange-dominated dispersion relation $\omega=h_0+k^2 l_\mathrm{ex}^2$ starting at the (Kittel) ferromagnetic resonance (FMR) frequency \begin{equation}\label{eq:fmr frequency}
\omega_K=|\omega_-|\overset{k\rightarrow 0}{=}  \left|\frac{1- \sqrt{1+4h_0\xi}}{2\xi}\right| \overset{\xi\ll 1}{\approx} |-h_0+h_0^2 \xi| \,,  
\end{equation}
where the weak influence of inertia is also recognizable.
Conversely, the higher frequency branch $\omega_+ $ describes intrinsic features of inertial dynamics occurring at frequencies larger than the following nutation resonance frequency:
\begin{equation}\label{eq:nutation resonance}
    \omega_N=\omega_+\overset{k\rightarrow 0}{=} \frac{1}{2\xi} \left(1+ \sqrt{1+4h_0\xi}\right)\overset{4h_0\xi \ll 1}{\approx} \frac{1}{\xi}+h_0 \,.
\end{equation}
It is also worth noting that both branches of the dispersion relation \eqref{eq:iLLG dispersion relation} give rise to the same group velocity (except for the sign):
\begin{equation}\label{eq:group velocity}
    % v_g(k)=\frac{\partial \omega}{\partial k} = \pm \frac{k l_\mathrm{ex}^2}{\xi\sqrt{\frac{1}{4\xi^2}+\frac{h_0+k^2 l_\mathrm{ex}^2}{\xi}}}, \,,
        v_g(k)=\gamma M_s \frac{\partial \omega}{\partial k} = \pm \frac{\gamma M_s\, 2 k l_\mathrm{ex}^2}{\sqrt{1+4(h_0+k^2 l_\mathrm{ex}^2)\xi}} \,,
\end{equation}
which remarkably approaches a finite value in the limit of short wavelength $k \rightarrow\infty$:
\begin{equation}\label{eq:limit group velocity}
    v_g(k\rightarrow \infty)=v_{g,\infty}=\frac{\gamma M_s\, l_\mathrm{ex}}{\sqrt{\xi}}=\frac{l_\mathrm{ex}}{\tau} \,.
\end{equation}
It is interesting to estimate the order of magnitude of the speed limit expressed by eq.\eqref{eq:limit group velocity}; by choosing $\gamma=2.211\times 10^5$ m$\,$A$^{-1}\,$s$^{-1}$,  $\tau= 0.653$ ps, $\mu_0 M_s=1.6$ T, $A=13$ pJ/m, one has $v_{g,\infty}\approx 5470$ m/s.

The dispersion relation and the group velocity expressed by eqs.\eqref{eq:iLLG dispersion relation}-\eqref{eq:group velocity} are depicted in fig.\ref{fig:dispersion relation} using the aforementioned material parameters of fcc Cobalt reported in Ref.\cite{Unikandanunni_PRL_2022}. We observe that neglecting inertial effects produces unlimited group velocity in the short wavelength limit (red line in bottom panel of fig.\ref{fig:dispersion relation} will diverge for vanishing wavelength, i.e $k\rightarrow\infty$). Conversely, the inertial spin waves related to both branches will approach the same limiting speed in the limit of large wavenumber (i.e. for $k\rightarrow\infty$, blue line will approach the value $v_{g,\infty}$ of eq.\eqref{eq:limit group velocity}).

Equation \eqref{eq:linear iLLG psi} provides additional insight when nonzero damping $\alpha\neq 0$ is considered. In fact, its nontrivial solutions obey the following equation:
\begin{equation}\label{eq:linear iLLG psi nontrivial}
    \xi \omega^2-\omega(1+i\alpha)-k^2 l_\mathrm{ex}^2-h_0 =0 \,.
\end{equation}
For a given frequency $\omega$, one can solve for the complex $k=k_\pm+i\delta k_\pm$ and obtain, in the limit $\alpha\ll 1$, the  wavenumber $k_\pm=\Re\{k\}$ and the spatial decay constant $\sigma_\pm=-\delta k_\pm=-\Im\{k\}$ of the plane wave $e^{-i k x}$ associated with the frequency $\omega$, respectively:
\begin{align}
    k_\pm&=\pm\frac{\sqrt{\xi\omega^2-\omega-h_0}}{l_\mathrm{ex}}+\mathcal{O}(\alpha^2) \label{eq:nutation k} \,, \\
    \delta k_\pm&=\mp \frac{\alpha\omega}{2 l_\mathrm{ex}\sqrt{\xi\omega^2-\omega-h_0}}+\mathcal{O}(\alpha^2) \,. \label{eq:spatial decay sigma} 
\end{align}
It is apparent that, at first order, $k_\pm$ does not depend on $\alpha$ whereas $\delta k_\pm$ is proportional to $\alpha$ as expected.
Furthermore, by solving eq.\eqref{eq:linear iLLG psi nontrivial} for small $\alpha\neq 0$, one can derive the time decay constant $\delta\omega_\pm$:
\begin{equation}\label{eq:time decay delta_omega}
    \delta\omega_\pm=\frac{\alpha}{2\xi}\left(1 \pm \frac{1}{\sqrt{4\xi k^2 l_\mathrm{ex}^2 + 4 h_0\xi +1}}\right) \,,
\end{equation}
which provides information on the temporal duration of the spin-wave as function of parameters. In particular, we remark that $2\delta\omega_+$ yields a simple estimate for the full-width half maximum (FWHM) linewidth $\Delta\omega_N$ of the power spectrum around the nutation resonance $\omega_N$ defined by eq.\eqref{eq:nutation resonance}, namely:
\begin{equation}\label{eq:nutation linewidth}
     \Delta\omega_N=2\delta\omega_+ \overset{k\rightarrow 0}{=} \frac{\alpha}{\xi}\left(1+ \frac{1}{\sqrt{4 h_0\xi +1}}\right) \overset{4 h_0\xi\ll 1}{\approx} %2\alpha\left(\frac{1}{\xi}-h_0\right) \,,
     \frac{2\alpha}{\xi}-2\alpha h_0 \,,
\end{equation}
where a weak dependence on $h_0$ (static external field and/or uniaxial anisotropy) appears. Analogously, one can derive the spectral linewidth $\Delta \omega_K$ of (Kittel) ferromagnetic resonance (FMR) in the presence of inertia:
\begin{equation}\label{eq:fmr linewidth}
     \Delta\omega_K=2\delta\omega_-\overset{k\rightarrow 0}{=} \frac{\alpha}{\xi}\left(1- \frac{1}{\sqrt{4 h_0\xi +1}}\right) \overset{4 h_0 \xi\ll 1}{\approx} 2\alpha h_0-6\alpha h_0^2 \xi \,.
\end{equation}

\begin{table}[t]
    \centering (a)
    \begin{tabular}{|l|c|}
    \hline
    Quantity     &  Equation\\
    \hline
    iSW dispersion relation $\omega_\pm(k)$ (units of $\gamma M_s$)  &  \eqref{eq:iLLG dispersion relation} \\
    FMR frequency $\omega_K$ vs inertia $\xi$ and static field $h_0$ & \eqref{eq:fmr frequency} \\
    FMR linewidth $\Delta\omega_K$ vs $\alpha,\xi,h_0$ & \eqref{eq:fmr linewidth} \\
    nutation resonance frequency $\omega_N$ vs $\xi,h_0$ & \eqref{eq:nutation resonance} \\
    nutation resonance linewidth $\Delta\omega_N$ vs $\alpha,\xi,h_0$ & \eqref{eq:nutation linewidth} \\
    group velocity $v_g(k)$ vs $\gamma,M_s,\xi,h_0,l_\mathrm{ex}$ & \eqref{eq:group velocity} \\
    limit group velocity $v_{g,\infty}=v_g(k\rightarrow\infty)$ & \eqref{eq:limit group velocity} \\
    iSW wavenumber $k_\pm$ vs $\xi,\omega,h_0,l_\mathrm{ex}$ & \eqref{eq:nutation k} \\
    iSW exp. spatial decay $\sigma_\pm=-\delta k_\pm$ vs $\alpha,\xi,\omega,h_0,l_\mathrm{ex}$ & \eqref{eq:spatial decay sigma} \\
    iSW exp. time decay $\delta \omega_\pm$ vs $\alpha,\xi,\omega,h_0,l_\mathrm{ex}$ & \eqref{eq:time decay delta_omega} \\
    
    \hline
    \end{tabular}
    (b)
    \begin{tabular}{|l|c|}
    \hline
    Parameter     &  Better choice \\
    \hline
    damping $\alpha$     &  low   \\
    exchange length $l_\mathrm{ex}=\sqrt{2A/(\mu_0 M_s^2)}$     &  high  \\
    inertia $\xi=(\gamma M_s \tau)^2$ & high   \\
    static field/anisotropy $h_0$ (weak dependence)& low  \\
\hline
    \end{tabular}
    \caption{(a) Summary of theoretical predictions and reference to equations in main text for quantities related with inertial spin waves (iSW) as function of material parameters. (b) Influence of parameters on excitation of short-wavelength inertial spin-waves with low spatial decay and long lifetime arising from inspection of eqs.\eqref{eq:nutation k}-\eqref{eq:nutation linewidth}}.
    \label{tab:parameters dependence}
\end{table}

The developed theory, whose main equations are summarized in table \ref{tab:parameters dependence}(a), can be instrumental for determining the conditions, in terms of material parameters and external excitation, suitable to excite spin-waves in the inertial regime. First, it is expected that, in order to excite spin-waves with low spatial decay and long lifetime, one needs material with very small damping. This is confirmed by eqs.\eqref{eq:nutation k}-\eqref{eq:nutation linewidth}. In addition, eq.\eqref{eq:spatial decay sigma} reveals less obvious inverse dependence of the spin-wave decay $\sigma_\pm=-\delta k_\pm$ on the exchange length $l_\mathrm{ex}=\sqrt{2A/(\mu_0 M_s^2)}$, which may favor materials with smaller saturation magnetization $M_s$. A summary of the influence of parameters suitable to produce inertial spin waves is provided in table \ref{tab:parameters dependence}(b). 
The completely new picture arising from inertial dynamics involves frequencies spanning the terahertz range above the nutation resonance frequency $\omega_N$ (see eq.\eqref{eq:nutation resonance}) that, for the chosen parameters of fig.\ref{fig:dispersion relation}, yields approximately $f_N\approx 1328$ GHz. For this reason, in the sequel we investigate spin wave dynamics occurring at frequency larger than the nutation resonance $f_N$. 

\begin{figure}[t]
    \centering
    \includegraphics[width=0.5\textwidth]{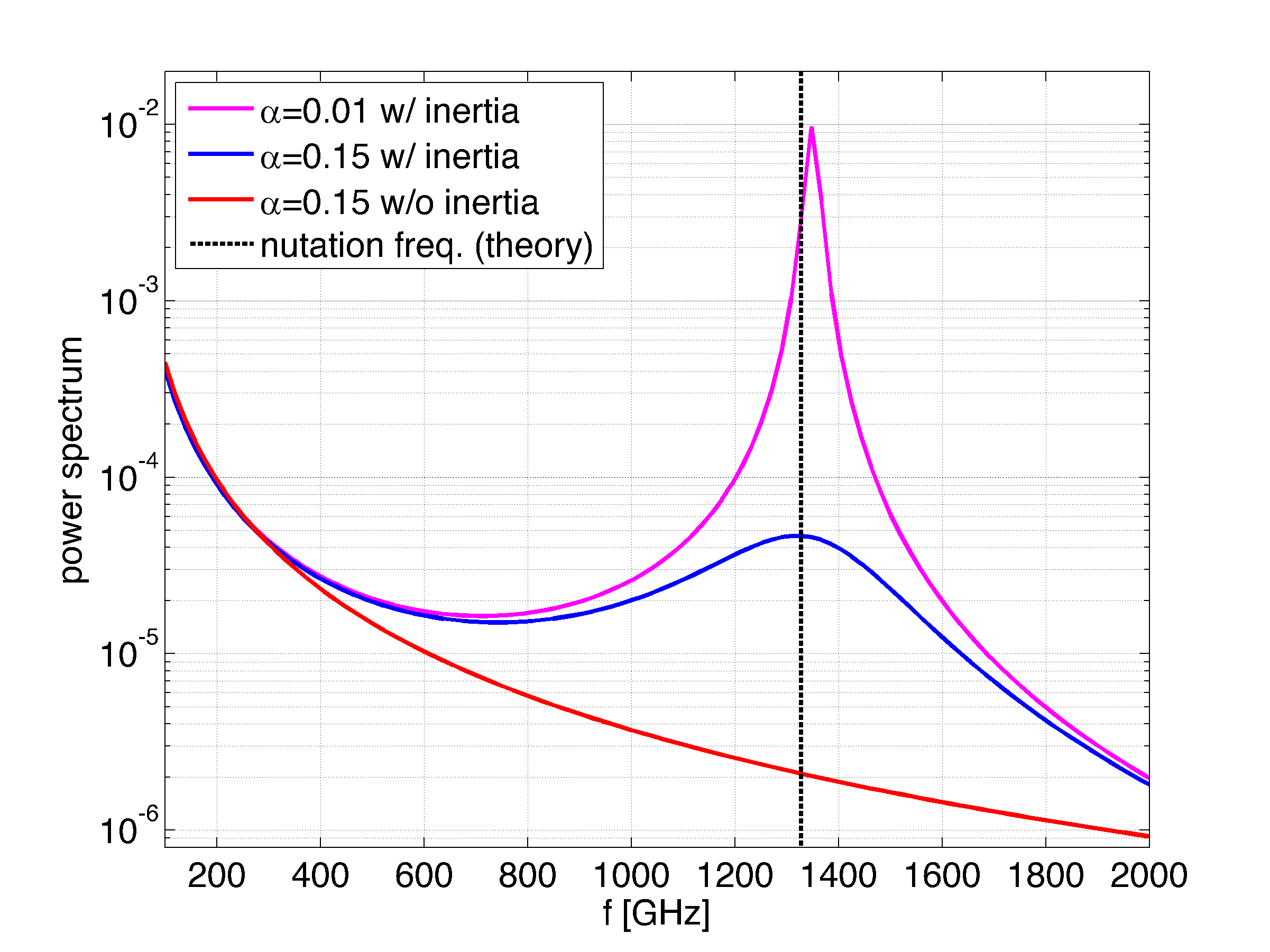}
    \caption{Power spectrum of magnetization computed by micromagnetic simulations according to eq.\eqref{eq:power spectrum}. Magenta and blue (red) lines refer to the presence (absence) of inertial effects in magnetization dynamics for different values of damping. The dashed black line refers to theoretical estimate \eqref{eq:nutation resonance} for nutation resonance frequency $f_N\approx 1328$ GHz. FWHM nutation linewidths predicted by eq.\eqref{eq:nutation linewidth} are approximately 26.5 GHz and 397 GHz for $\alpha=0.01$ and $\alpha=0.15$, respectively, in agreement with the results of micromagnetic simulations (27.1 GHz and 401 GHz).}
    \label{fig:power spectrum}
\end{figure}

\section{Micromagnetic simulations}

\begin{figure*}[t]
    \centering 
    \includegraphics[width=0.49\textwidth]{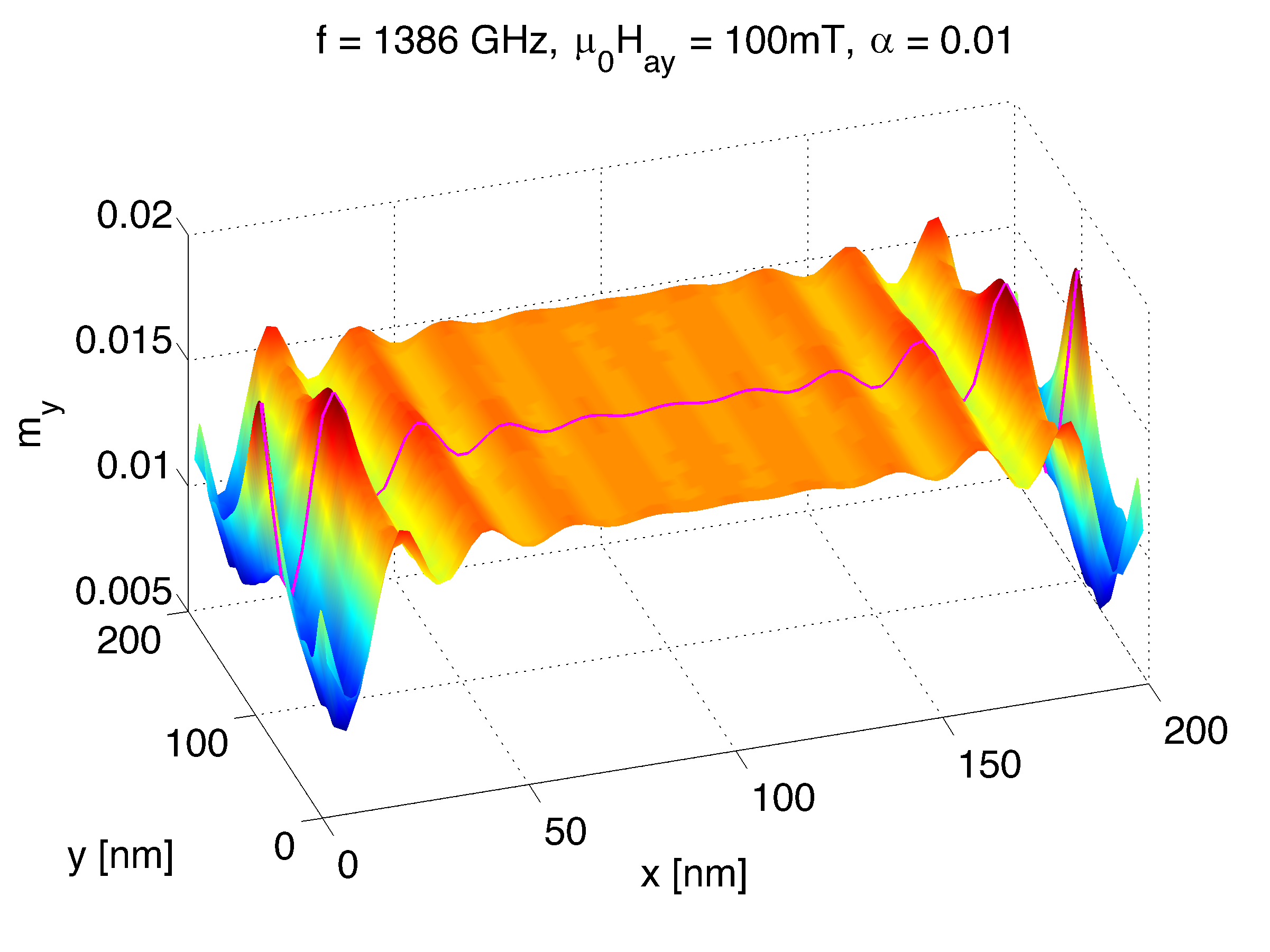}
    \includegraphics[width=0.49\textwidth]{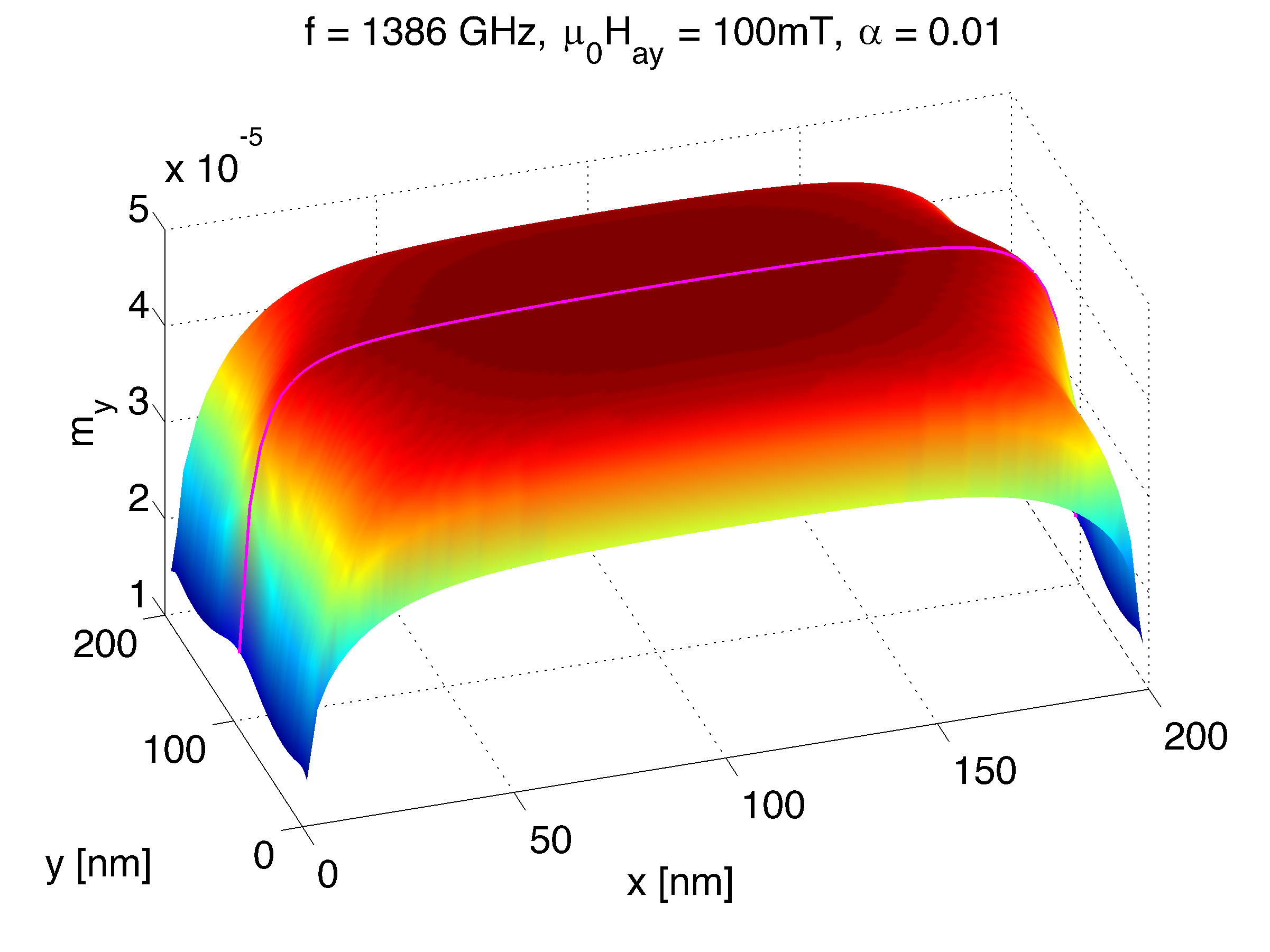}
    \caption{Snapshot at $t=0$ of steady-state in-plane ac magnetization response $m_y$ at driving frequency $f=1386$ GHz. The color code represents the value of $m_y$ at each spatial location ranging from minimum (blue) to maximum (red). Left panel refers to inertial LLG dynamics while right panel refers to classical LLG dynamics. Magenta solid lines refer to $m_y(x)$ sampled in the middle of the nanodot at $y=100$nm.}
    \label{fig:magnetization response}
\end{figure*}

In order to assess the excitation of ultra-short wavelength inertial spin waves, we perform two independent studies, one in the frequency domain and the other in the time domain, involving a square $200\times 200\times 5$ nm$^3$ thin-film nanodot, with the same material parameters as those of fig.\ref{fig:dispersion relation}, saturated along the $x$ axis by a static field $\mu_0 H_{ax}=100$ mT. 
% First, by using the dispersion relation \eqref{eq:iLLG dispersion relation}, we infer that inertial spin-waves with wavelength $2\pi/k\approx 20$nm are associated with a frequency around $1386$ GHz. Then, 
In the former situation, we investigate ac steady-state magnetization dynamics for the aforementioned nanodot driven by a spatially-uniform linearly-polarized ac field  transverse to the equilibrium magnetization with frequency $f=1386$ GHz and amplitude $\mu_0 H_{ay}=100$ mT.  
The magnetization is initially in the remanent equilibrium configuration under the static external magnetic field, which is mainly oriented along the $x$ axis with deviations located close to the thin-film edges parallel to the $y$ axis.

\subsection{Frequency-domain study}

The ac forced magnetization dynamics can be conveniently studied by solving the linearized iLLG in the frequency domain and determining the frequency response and the power spectrum.  Here we use a finite-difference frequency-domain large-scale micromagnetic solver\cite{dAquino_JAP_2023} based on suitable operator formalism\cite{daquino_novel_2009}, appropriately extended in order to include inertial effects and implemented in the numerical code MaGICo\cite{MaGICo}. The output of the code is the steady-state ac magnetization response $\delta\hat{\bm m}(\bm r)$ (such that $\delta\bm m(\bm r,t)=\mathcal{R}\{\delta\hat{\bm m}(\bm r) \exp{(i\omega t)}\}$) computed for desired values of the external ac field frequency $\omega$ (see ref.\cite{dAquino_JAP_2023} for further details). For each given $\omega$, the ac power spectrum of magnetization is then computed as\cite{dAquino_JAP_2023}:
 \begin{equation}\label{eq:power spectrum}
     p(\omega)=\frac{1}{V}\int_\Omega \frac{|\delta\hat{\bm m}(\bm r)|^2}{2}\, dV \approx \frac{1}{N}\sum_{j=1}^N \frac{|\delta\hat{\bm m}_j|^2}{2}   \,,
 \end{equation}
 where $V$ is the volume of the region $\Omega$ occupied by the magnetic body, discretized with $N$ computational prism cells of identical volume (in the present case we use $2.5\times 2.5\times 5$ nm$^3$ cells). 

The ac power spectrum of magnetization, computed according to eq.\eqref{eq:power spectrum} in the terahertz frequency range, is reported in fig.\ref{fig:power spectrum} as function of the damping $\alpha$. It is apparent that there is a spectral peak owing to nutation resonance close to the theoretical value $f_N\approx 1328$ GHz (dashed vertical black line in fig.\ref{fig:power spectrum}), while no such peak appears in the classical precessional ac-driven LLG dynamics. FWHM nutation linewidths predicted by eq.\eqref{eq:nutation linewidth}  (26.5 GHz and 397 GHz for $\alpha=0.01$ and $\alpha=0.15$, respectively) are also in agreement with those computed from micromagnetic simulations (27.1 GHz and 401 GHz, respectively). We have used material parameters for fcc Cobalt such as those reported in a recent experimental work\cite{Unikandanunni_PRL_2022}, where a quite large damping $\alpha=0.15$ was estimated from the measurements of nutation frequency response.

With that in mind, we follow the predictions of the developed theory concerning short-wavelength spin waves and, therefore, we explore the magnetization response at frequency $f=1386$ GHz. 
In this respect, eq.\eqref{eq:nutation k} for the above frequency predicts a wavelength of 20 nm for the excited spin-waves. However, eq.\eqref{eq:spatial decay sigma} particularized with $\alpha=0.15$ yields a spatial decay constant $\sigma_+$ such that the excited spin-waves would be exponentially attenuated by more than two orders of magnitude within a distance $\sim 5/\sigma_+$ which is less than 10 nm, meaning that nutation waves would not be observable in this condition. Thus, according to eq.\eqref{eq:spatial decay sigma}, in order to have spin waves that are 20 nm long and can extend for hundred nanometers, one needs materials with damping in the order of $\alpha\sim 10^{-2}$. Moreover, as predicted by eq.\eqref{eq:spatial decay sigma} and reported in table \ref{tab:parameters dependence}, the detrimental effect of large damping could be mitigated, in principle, by using materials with larger exchange length (i.e. smaller saturation magnetization). In this respect, Cobalt seems not to be the best choice, since it has notably quite small value of $l_\mathrm{ex}\approx 3.57$ nm due to its large saturation magnetization such that $\mu_0 M_s=1.6$ T. 

The result of micromagnetic simulations with $\alpha=0.01$ is reported in fig.\ref{fig:magnetization response}. One can clearly see (left panel of fig.\ref{fig:magnetization response}) that inertial spin wave dynamics driven at frequency $f>f_N$ involves ultra-short nutation spin waves with wavelength around 20 nm (one can count roughly ten oscillation periods along the $x$ direction), which is in excellent agreement with the dispersion relation eq.\eqref{eq:iLLG dispersion relation} (see fig.\ref{fig:dispersion relation}). As expected, the excitation frequency only matches the upper branch of eq.\eqref{eq:iLLG dispersion relation}. On the other hand, when inertia is not considered (right panel of fig.\ref{fig:magnetization response}), the oscillation amplitude is two orders of magnitude smaller than in the inertial case, and short spin waves are not excited anymore. 

The time-evolution of the ac steady-state magnetization $\delta\bm m(\bm r,t)=\mathcal{R}\{\delta\hat{\bm m}(\bm r) \exp{(i\omega t)}\}$ basically consists of a superposition of two oscillations, the former having  almost spatially-uniform profile and the other with wavelength around 20 nm. One can interpret the former as describing spatially-uniform magnetization nutation and the second being the spatially-inhomogeneous magnetization nutation ascribed to the excitation of inertial spin waves. This can be also inferred observing the comparison between the spatial patterns of magnetization responses in the presence and absence of inertial effects, sampled at time $t=0$ and reported in the two panels of fig.\ref{fig:magnetization response}. We immediately remark that the maximum amplitude of magnetization is significantly different in the two situations. This occurs since, in the inertial case, the external excitation at frequency 1386 GHz is close to the resonant peak (see also fig.\ref{fig:power spectrum}) that enhances the steady-state response of the system, compared with the situation of classical LLG precessional dynamics where there is no such resonance. Apart from the aforementioned situation, one can grasp the similarity between the long wavelength magnetization profiles.   
\begin{figure}[t]
    \centering
    \includegraphics[width=0.45\textwidth]{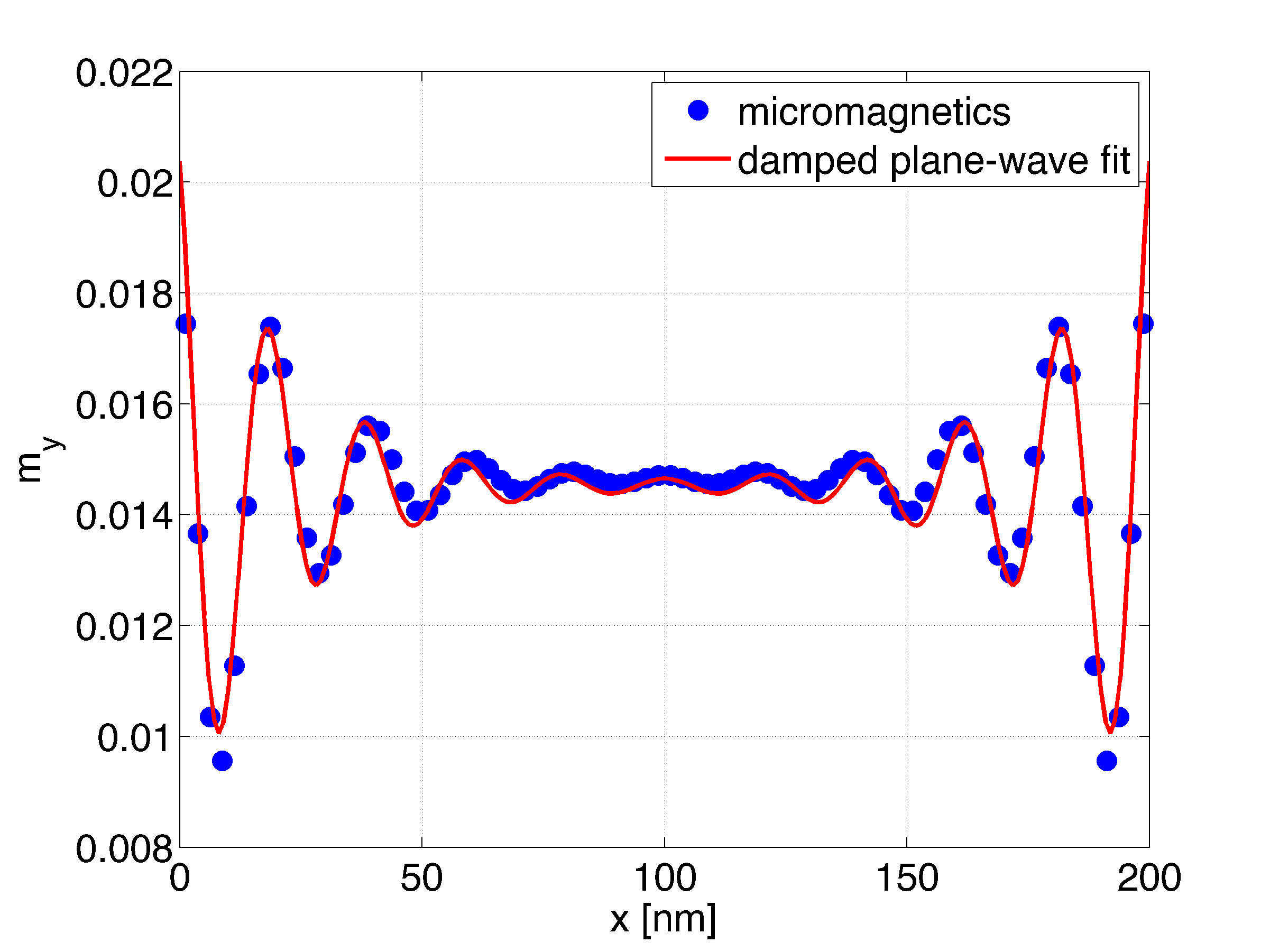}
    \caption{Spatial profile along $x$ axis of magnetization in-plane component $m_y$ sampled at $t=0$ along the middle line $y=L/2=100$ nm. Dots refer to micromagnetic simulation while solid line is a fit to the two damped plane wave ansatz eq.\eqref{eq:plane wave ansatz}.}
    \label{fig:pwfit_my}
\end{figure}
As far as the quantitative comparison of spin-wave amplitudes associated with spatially-uniform and nonuniform nutations is concerned, we have evaluated the relative difference between the power spectrum computed according to Eq.\eqref{eq:power spectrum} from full micromagnetic simulations with that computed using Eq.\eqref{eq:power spectrum} only with the spatially-averaged magnetization $\delta\hat{\bm m}_\text{avg}=\frac{1}{N}\sum_{j=1}^N \delta\hat{\bm m}_j$, namely $p_\text{avg}(\omega)=\delta \hat{\bm m}_\text{avg}^2/2$. The relative difference $(p(\omega)-p_\text{avg}(\omega))/p_\text{avg}(\omega)$ is below one percent in the whole considered frequency range. Despite such a small relative weight in the power spectrum, the ultra-short spin-wave oscillation is clearly visible in figures \ref{fig:magnetization response} and \ref{fig:pwfit_my}.

In order to compare the theoretical prediction obtained in the ideal case of infinite thin-film with the results of simulations for confined structures, we have characterized the short inertial spin waves by fitting the spatial profile of the ac steady-state in-plane oscillation $m_y$ (sampled along $x$ in the middle line at $y=100$nm) with the following two damped plane wave ansatz:
\begin{equation}\label{eq:plane wave ansatz}
    \psi(x)=a+b\frac{e^{-\sigma x}\cos(k x+\beta)+e^{-\sigma (x-L)}\cos(k (x-L)+\beta)}{2} \,,
\end{equation}
where we have fixed $k=2\pi/20$ rad/nm, $L=200$ nm. The result is reported in table \ref{tab:pw fitting param} and the comparison with simulations is reported in fig.\ref{fig:pwfit_my}, showing nice agreement with the assumption of plane wave profile. 
\begin{table}[h]
    \centering
    \begin{tabular}{|c|c|c|}
    \hline
    Parameter     &  Value & confidence intervals\\
    \hline
    $a$     &  0.01453  & (0.0144, 0.01465) \\
    $b$     &  0.01305 & (0.01131, 0.01478) \\
    $\beta$ & 0.4592  &  (0.3785, 0.54)  \\
    $\sigma$ & 0.04541 & (0.03758, 0.05324) \\
\hline
    \end{tabular}
    \caption{Values of the fitting parameters for eq.\eqref{eq:plane wave ansatz}. Notice that $k=2\pi/20$ nm and $L=200$ nm.}
    \label{tab:pw fitting param}
\end{table}

We observe that the spatial decay constant $\sigma\approx 0.045$ extracted from the fitting  is in good agreement with the value $\sigma_+\approx 0.0374$ predicted by eq.\eqref{eq:spatial decay sigma} despite the latter is based on the assumption of spatially-uniform equilibrium $\bm m_0$, whereas the actual equilibrium magnetization close to the boundaries of the nanodot has slightly different orientation compared to that in the center due to the magnetostatic field created by the magnetic charges arising from the confinement\cite{Gubbiotti2004}. In addition, as mentioned before, we notice that the nonzero offset $a=0.01453$ corresponds to a significant spatially-uniform component of $m_y(x)$, superimposed to the plane wave mode of maximum amplitude $b=0.01305$, which is also associated with spatially-uniform magnetization nutation at the same frequency 1386 GHz.

\subsection{Time-domain study}

The results outlined in the previous section reveal the possibility to excite ultra-short spin waves by using ac terahertz excitation. However, full understanding of the profoundly different nature of inertial magnetization dynamics compared to the classical precessional dynamics can be achieved by complementing frequency-domain calculations with time-domain analysis of transient magnetization dynamics. This difference is apparent in the mathematical structure of eq.\eqref{eq:iLLG} compared with the same equation with $\xi=0$. The torque proportional to the second-order derivative transforms the classical LLG equation into a wave-like equation with hyperbolic mathematical nature.
%
% can be appreciated if one observes the modified mathematical character of eq.\eqref{eq:iLLG} in the presence of inertial effects. In fact, in this situation, eq.\eqref{eq:iLLG} includes a (small) torque proportional to the second-order time-derivative of magnetization, which transforms the classical LLG equation into a wave-like equation with hyperbolic mathematical nature. 

This means that inertia leads to wave propagation phenomena with finite speed. In this respect, we investigate the transient magnetization dynamics triggered by the action of a terahertz field step when the initial magnetic state is  the static remanent equilibrium configuration. 

To this end, we perform full micromagnetic simulations of eq.\eqref{eq:iLLG}  using the finite-difference numerical code MaGICo\cite{daquino2005geometrical,MaGICo} which is able to perform fast and large-scale time-integration of iLLG dynamics. Thus, we integrate iLLG equation \eqref{eq:iLLG} with a time-step of 25 fs, a $2.5\times 2.5\times 5$ nm$^3$ computational cell, and record the space configuration of magnetization at each time-step. The applied field is a spatially-uniform sine wave step (turned on at $t=0$) along the $y$ axis transverse to the equilibrium configuration with the same amplitude $\mu_0 H_{ay}=100$ mT and frequency $f=1386$ GHz as in the frequency-domain study. Despite using the same excitation, the present situation offers the possibility to look at the propagation of inertial short-wavelength waves along the magnetic thin-film evidencing the finite time delay. 
\begin{figure}[t]
    \centering
    \includegraphics[width=0.47\textwidth]{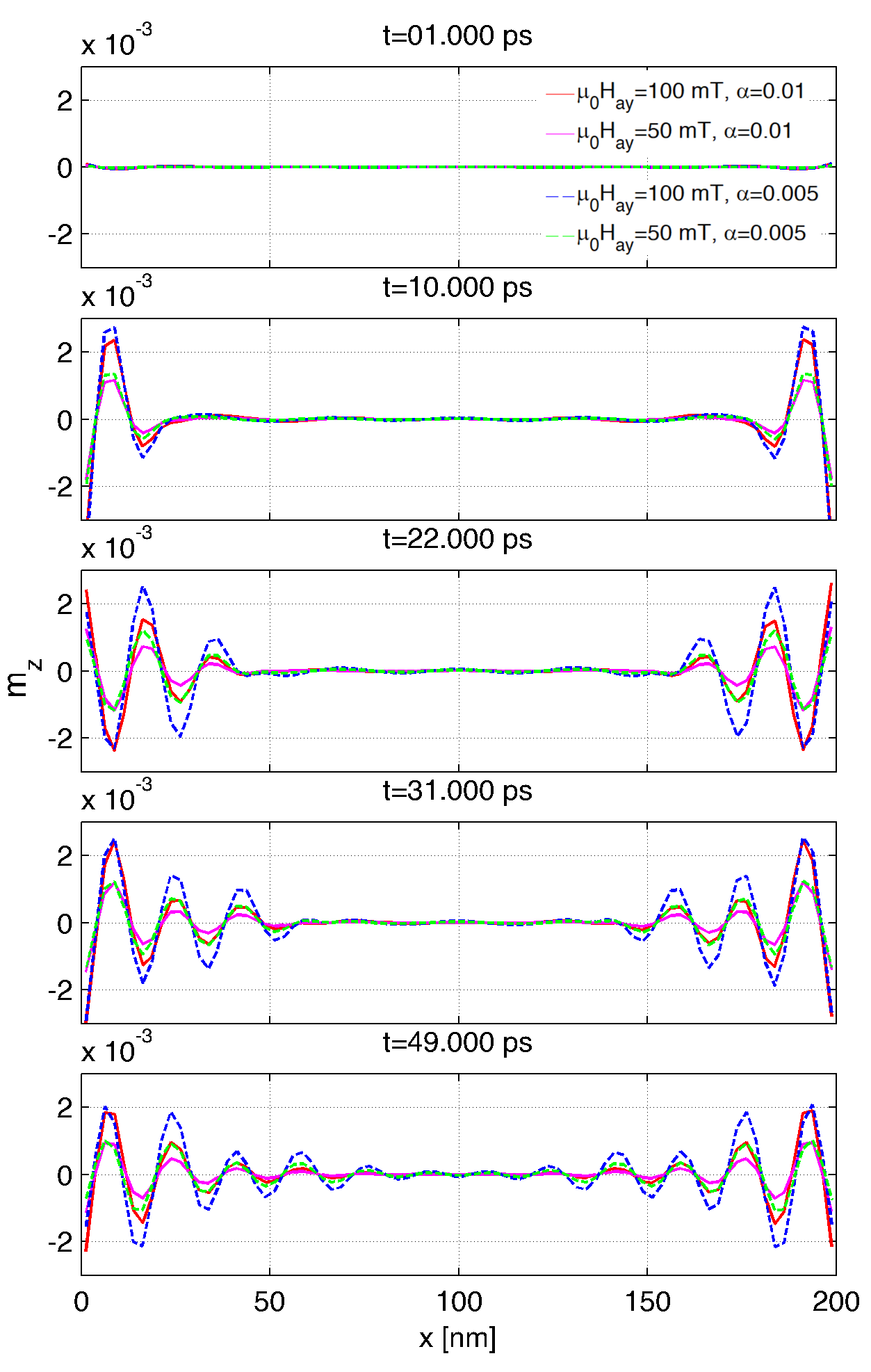}
    \caption{Spatial profiles  along $x$ axis of short-wavelength magnetization out-of-plane component $m_z$ as function of ac field amplitude and damping, obtained by FFT high-pass filtering, sampled at different time instants.}
    \label{fig:propagation}
\end{figure}
We remark that the choice of a spatially-uniform excitation field is related with the possibility to realize this experiment with realistic laser sources that create a spot much larger than the dimension of the considered magnetic nanodot. The spatial uniformity of the applied field step obviously produces the transient excitation of a plethora of spin wave modes with very different wavelengths, which makes difficult the direct inspection of the inertial spin wave propagation. For this reason, also based on the agreement with the plane wave nature of inertial spin waves demonstrated by the frequency-domain simulations, we apply spatial high-pass Fast Fourier Transform (FFT) filter (the cutoff wavenumber is $0.2$ rad/nm) to capture the evolution of short-wavelength spin waves. The so-obtained short wavelength magnetization patterns of the resulting out-of plane magnetization sampled along the middle line of the thin-film square at different time instants are reported
%\footnote{a movie of the whole magnetization dynamics has been realized} 
in fig.\ref{fig:propagation}.

The initial equilibrium magnetization configuration lies in the sample plane with more pronounced deviations from the $x$ orientation localized in the region close to the edges perpendicular to the static field (i.e. those at $x=0$ and $x=200$ nm). Then, one can clearly see from fig.\ref{fig:propagation} that, when the sine wave step is applied, the magnetization response originates from the edges and gives rise to plane waves with wavelength around 20 nm propagating along the thin-film towards its center. The simulated experiment is repeated for two values of damping $\alpha=0.005,0.01$ to investigate spin wave spatial decay and for two ac field amplitudes $50,100$ mT in order to check the linearity of the response. In this respect, on one hand one can clearly see in fig.\ref{fig:propagation} that, for a given field amplitude, lower damping implies smaller spatial decay. On the other hand, it happens that doubling the ac field amplitude produces a magnetization response with doubled amplitude, which assesses the linear nature of inertial spin-wave dynamics despite the application of terahertz ac fields with moderately large amplitude. Although the effect of damping produces the decay of the oscillation, it is also apparent that the two wavefronts of the forward and backward wavepackets approach the center of the square in around 49 picoseconds, which allows us to roughly estimate the group velocity of the inertial spin waves around 2040 m/s, which is in striking agreement with the theoretical prediction $v_g\approx 2080$ m/s given by eq.\eqref{eq:group velocity} for $k=2\pi/20$ rad/nm. This value of speed amounts to slightly less than half the ultimate speed limit $v_{g,\infty}\approx 5470$ m/s predicted by eq.\eqref{eq:limit group velocity}.

\begin{figure}[t]
    \centering
    \includegraphics[width=0.5\textwidth]{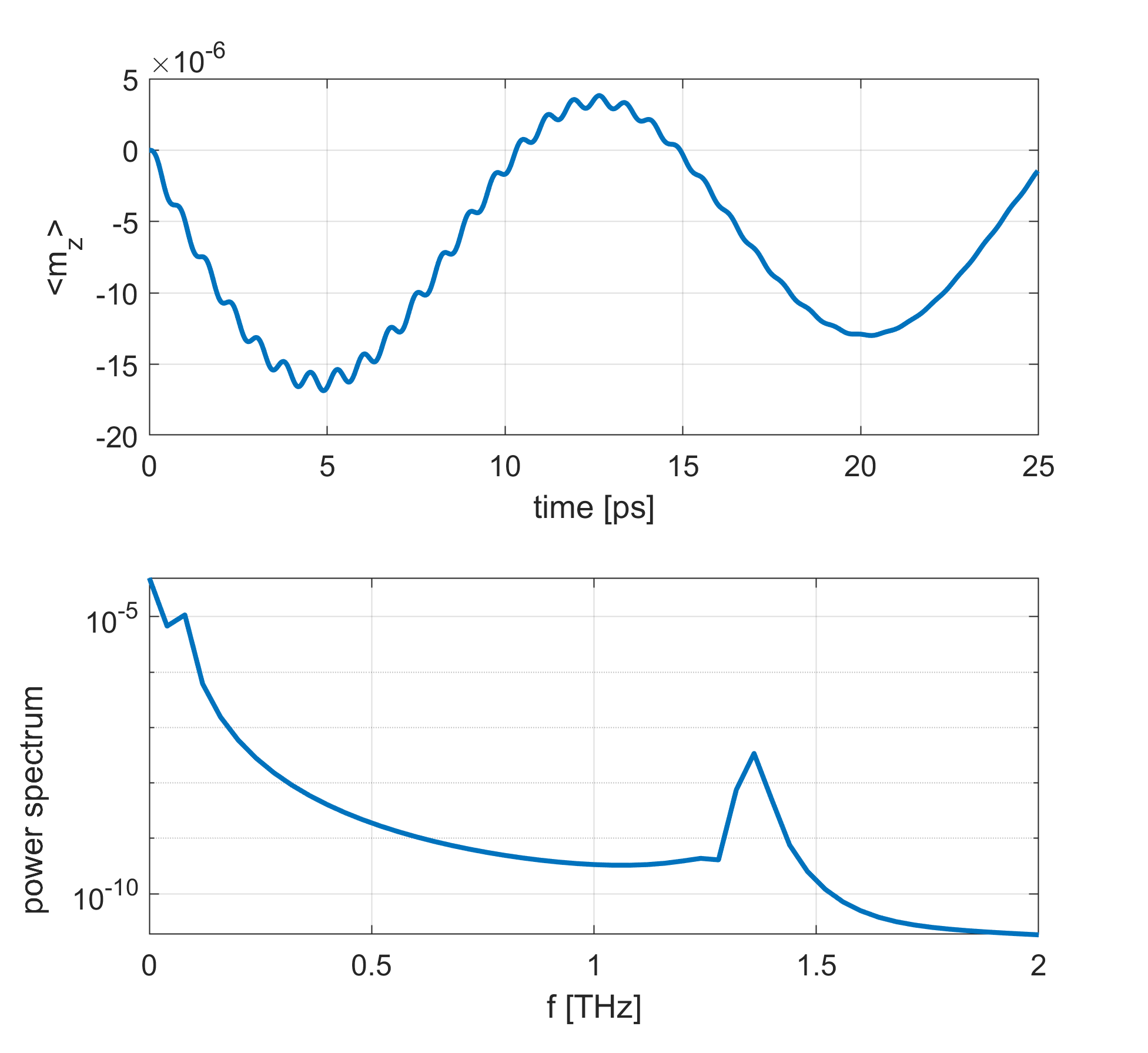}
    \caption{Free magnetization response starting from periodically-modulated initial state with spatial period 20 nm. The upper panel reports the time-evolution of the spatially-averaged out-of-plane component $\langle m_z \rangle$, the lower panel reports the power spectrum (estimated by the periodogram, i.e. the squared amplitude of the FFT) of $\langle m_z \rangle$. Spectral peaks at about 60GHz and 1380 GHz are apparent.}
    \label{fig:free evolution}
\end{figure}

Finally, we perform an additional investigation of transient magnetization dynamics starting from an initial magnetization deviating from the saturated state by a periodically modulated wave, which can be instrumental to check the natural oscillation frequency associated with plane wave perturbations of given wavenumber. The outcome of the simulation starting from the initial state $\bm m=(1,\epsilon \sin(k x), 0)/\sqrt{1+\epsilon^2\sin^2(k x)}$, with $k=2\pi/20$ rad/nm and $\epsilon=10^{-2}$) is reported in fig.\ref{fig:free evolution}. One can clearly see in the upper panel that magnetization exhibits the composition of two oscillation, the former at low frequency (around 60 GHz, with period $\sim 16$ ps) and the latter superimposed to the former at much higher frequency $1380$ GHz (notice that the FFT frequency resolution $\pm 40$ GHz is quite coarse due to the short time record to analyze). The low frequency oscillation can be ascribed to classical precessional dynamics, as it can be inferred by looking at the lower branch of the dispersion relation \eqref{eq:iLLG dispersion relation} (see also fig.\ref{fig:dispersion relation}), which yields a frequency $|\omega_-| \rightarrow 57$ GHz (59 GHz using the classical exchange spin-wave dispersion relation $\omega=h_0+l_\mathrm{ex}^2 k^2$). 

On the other hand, one can see that the high frequency oscillation is associated with inertial nutation spin waves with the same wavelength 20 nm. This is also consistent with the upper branch of the dispersion relation $\omega_+$ (see fig.\ref{fig:dispersion relation}). As it can be seen in the upper panel of fig.\ref{fig:free evolution}, these spin waves have much shorter lifetime which is apparently around 15 ps, after which they disappear due to damping. The spin wave lifetime (corresponding to exponential attenuation by more than two orders of magnitude) predicted by the developed theory (see eq.\eqref{eq:time decay delta_omega}) is $5/(\delta\omega \,\gamma M_s) \approx 19$ ps. This last simulation confirms the theoretical predictions and the results of the previously outlined numerical studies.

\section{Conclusions}

In this work, we have investigated the possibility to excite ultra-short inertial spin with behavior that significantly deviates from that of the classical exchange spin-waves. A theoretical approach has been developed to determine the dispersion relation, the spatial decay and the lifetime of inertial spin waves as function of material and excitation parameters. It turns out that ultra-short (20 nm) inertial spin wave propagation in Cobalt films occurs at terahertz frequencies and admits a limiting speed in the order of 5500 m/s. Micromagnetic simulations both in frequency and time domains confirm the theoretical predictions concerning the possibility to excite finite speed propagation of such short waves in confined ferromagnetic nanodots by using terahertz ac fields with quite large amplitudes ($\sim 100$ mT) such as those achievable with state-of-the art terahertz experimental setups. For these reasons, this study can be instrumental to stimulate and guide the design of experiments aiming to the observation of inertial spin wave dynamics.

\begin{acknowledgements}
M.d'A., S.P., M.P., S.B. and C.S. acknowledge support from the Italian Ministry of University and Research, PRIN2020 funding program, grant number 2020PY8KTC.
% This work has been supported by the Italian MIUR-PRIN 2020 project MAGNUT with grant number 2020PY8KTC. 
\end{acknowledgements}

\bibliography{inertial_SW.bib}

\newpage

% \appendix

% \section{Supplemental material}

\end{document}